\documentclass[a4paper, oneside, twocolumn, notitlepage, 10pt]{extarticle_ecoc}
\usepackage{ecoc}
\newcommand{\yhd}{\bm{w}}
\usepackage{bm}
\usepackage{pgfplots}
\usepackage{booktabs}
\usepackage[ruled,vlined,algo2e]{algorithm2e}
\begin{document}
\selectlanguage{english}    %

\title{Improved Chase--Pyndiah Decoding for Product Codes \\ with Scaled Messages}%

\author{
    Sisi Miao, Mert Birincio\v{g}lu, Laurent Schmalen
}

\maketitle                  %

\begin{strip}
    \begin{author_descr}
    
        Communications Engineering Lab, Karlsruhe Institute of Technology,  Germany \textcolor{blue}{{\uline{sisi.miao@kit.edu}}}
        
    \end{author_descr}
\end{strip}

\renewcommand\footnotemark{}
\renewcommand\footnoterule{}

\begin{strip}
    \begin{ecoc_abstract}
We propose an enhanced Chase--Pyndiah decoder that scales extrinsic messages based on decoder confidence of the component decoder, achieving a 0.1\,dB gain over the original with negligible complexity increase. \textcopyright~2026 The Author(s).
       
    \end{ecoc_abstract}
\end{strip}

\section{Introduction}

Product codes (PCs) \cite{elias1955coding} and PC-like codes are widely used in high-throughput communication systems for forward error correction (FEC) due to their efficient iterative decoding and strong performance. The Chase--Pyndiah algorithm \cite{PyndiahTPC} is a widely-used soft-decision decoding algorithm for PCs. However, its performance degrades with a limited number of test patterns and when extrinsic messages are approximated under complexity constraints. Various works have been devoted to addressing these issues. One direction is by enhancing extrinsic message processing. In \cite{JanzSOCS}, an extrinsic message update rule based on the covered space is derived, resulting in more accurate messages. The proposed decoder is termed SOCS($\mathcal{B}_t(\mathcal{T})$ together with a simplified version of it, termed SOCS($\beta$) decoder. The work of \cite{StrasshoferGMI} proposes to compute extrinsic messages based on the individual list of candidate codewords for each bit. However, both approaches significantly increase the computational complexity due to modifications of the message computation. Alternatively, performance can be improved by scaling or suppressing extrinsic messages based on decoder confidence, without modifying their computation. Confidence estimation has been explored via transformer-based models~\cite{ArtemasovTransformer}, mappings from destructive Euclidean distance~\cite{ShenLogistic}, and top-1/top-2 candidate codewords~\cite{ArtemasovTransformer}.

In this work, we propose a simple logistic regression-based confidence estimator that achieves performance close to the works in \cite{JanzSOCS,StrasshoferGMI,ArtemasovTransformer} with significantly lower complexity.

\textit{Notation}: We use boldface uppercase letters to denote matrices, e.g., $\bm{Y}$ and boldface lowercase letters to denote vectors, e.g., $\bm{y}$. The $i$-th row of $\bm{Y}$ is denoted as $\bm{y}_i$ and the $j$-th element of $\bm{y}_i$ is denoted as $y_{i,j}$. We define a function  $\mu(x) = 1-2x$ and $\psi(x) = (1-\text{sign}(x))/2$. Both $\mu(x)$ and $\psi(x)$ apply to matrices element-wise.

\section{Preliminaries}

Consider a PC \(\mathcal{C}_{\text{PC}}\) constructed from an \([n,k,d]\) BCH component code \(\mathcal{C}_{\text{BCH}}\). A codeword of \(\mathcal{C}_{\text{PC}}\) is an \(n\times n\) array whose rows and columns belong to \(\mathcal{C}_{\text{BCH}}\). The code has length \(n^2\), dimension \(k^2\), and rate \(R=(k/n)^2\).

Binary phase-shift keying is used over a binary-input AWGN channel with noise variance \(\sigma^2\). Let \(\bm{B}\in\{0,1\}^{n\times n}\) denote the transmitted PC codeword. The received symbols are \(y_{i,j}=\mu(b_{i,j})+n_{i,j}\), where \(n_{i,j}\sim\mathcal{N}(0,\sigma^2)\), yielding \(\bm{Y}\).

The Chase--Pyndiah decoder then operates by iteratively updating the log likelihood-ratio for all the bits. In each decoding iteration, Chase decoding \cite{Chase}, followed by an extrinsic information update, is applied first to all rows and then to all columns, corresponding to two half-iterations.

For a single row or column, let $\bm{r} \in \mathbb{R}^n$ denote its soft information, with the hard-decision vector $\psi(\bm{r}) =: \yhd \in \{0,1\}^n$. The absolute values $|r_i|$ of $\bm{r}$ are sorted, yielding the $p$ least reliable bit position with the smallest $|r_i|$, denoted by set $\mathcal{P}$. 
For the commonly used Chase-II decoding, a set of \textit{flipping patterns} 
$\mathcal{F} = \{\bm{f}_1, \bm{f}_2, \ldots, \bm{f}_{2^p}\}$ is constructed 
such that, for each $\bm{f}_i$, the bits at the positions in $\mathcal{P}$ 
correspond to the binary representation of the decimal number $i-1$, 
and are zero elsewhere.
Then, the \textit{test pattern} set $\mathcal{T} = \{\bm{t}_1,\bm{t}_2,\ldots, \bm{t}_{2^p}\}$ is defined by $\bm{t}_{i} = \bm{w}\oplus\bm{f}_i\in \{0,1\}^n$. Bounded-distance decoding (BDD) \cite{rothbook} is performed for all $\bm{t}_i$, resulting in a set of candidate codewords $\Omega = \{\bm{c}_1,\bm{c}_2,\ldots,\bm{c}_{|\Omega|}\}$ with $|\Omega|\leq 2^p$ as some $\bm{t}_{i}$ may result in a BDD failure. For each $\bm{c}_i$, $i\in \{1,\ldots, |\Omega|\}$, we compute a Euclidean distance metric $d^{\text{E}}_i=||\bm{r} -\mu(\bm{c}_i)||^2$ and the Chase decoding decision \(\bm{d}=\arg\min_{\bm{c}_i\in\Omega}\|\bm{r}-\mu(\bm{c}_i)\|^2\). 

Then, the extrinsic message $\bm{l}\in \mathbb{R}^n$ is computed. For the $j$-th bit, we try to find a competing codeword $\bar{\bm{d}} = \arg\min_{\bm{c}_i\in\Omega, c_{i,j}\neq d_j}\|\bm{r}-\mu(\bm{c}_i)\|^2$. Then,
\begin{equation}\label{eq:extrinsicInfo}
    l_j \!=\!\! \begin{cases}
        \mu(d_j) \frac{(||\bm{r}-\mu(\bm{\bar{d}})||^2 - ||\bm{r}-\mu(\bm{d})||^2 )}{4} - r_j & \!\!\!\text{if }\bar{\bm{d}} \text{ exists}\\
        \mu(d_j)\cdot \beta &\!\!\!\! \text{ otherwise}
    \end{cases}
\end{equation}
where $\beta$ is a half-iteration dependent optimizable parameter. 

After computing $\bm{l}_i$ for all the component codewords in one half iteration, the value of $\bm{L}\in \mathbb{R}^{n\times n}$ is normalized such that the mean of $|\bm{L}|$ is $1$. Then, we update the soft information $\bm{R} = \alpha \bm{L} + \bm{Y}$, which is used as input for the next half iteration. Here, $\alpha$ is another optimizable parameter which is half-iteration dependent. The process is repeated for \(I\) iterations (Alg.~\ref{algo:decoder}). For small \(I\), a high error floor is observed; thus, the soft-decision decoding is followed by two iterations of iterative bounded-distance decoding (iBDD).

\begin{algorithm2e}[t]\footnotesize
\caption{Original and NN-assisted Chase--Pyndiah decoding for PCs}
\label{algo:decoder}
\KwIn{$\bm{Y} \in \mathbb{R}^{n\times n}$}
$\bm{L} \gets \bm{0}^{n\times n}$, \quad $\bm{R} \gets \bm{Y}$\;

\For{$\ell = 1,2,\dots,2I$}{
    \For{$i = 1,2,\dots,n$}{
        $(\Omega,\bm{d}) \gets$ Chase decoding of $\bm{r}_i$\;
        \If{\textbf{NN-assistend}}{
        \textbf{flag}$_{i}\gets$ NN eval($\Omega$,$\bm{d}$)\;
        }
        \For{$j = 1,2,\dots,n$}{
            search for $\bar{\bm d}$ in $\{\bm{c}\in\Omega: c_j\neq d_j\}$\;
            compute $l_{i,j}$ using \eqref{eq:extrinsicInfo} and $\beta_{\ell}$\;
        }
        
    }
    Normalize $\bm{L}$\;
    \For{$i = 1,2,\dots,n$}{
    \If{\textbf{NN-assisted} and \textbf{flag}$_{i}$}{
        $\bm{l}_i\gets \gamma_{\ell} \bm{l}_i$\;
        }
     }
    $\bm{R} \gets \bm{Y} + \alpha_{\ell} \bm{L}$\;
    $\bm{Y} \gets \bm{Y}^{\mathsf T}$,\quad
    $\bm{R} \gets \bm{R}^{\mathsf T}$,\quad
    $\bm{L} \gets \bm{L}^{\mathsf T}$\;
}
\KwOut{$\mu(\bm{R}) \in \{0,1\}^{n\times n}$}
\end{algorithm2e}
\vspace{-0.5em}

\section{Related Works}
For sufficiently large \(p\), the Chase decoder becomes a maximum-likelihood (ML) decoder. In practice, however, relatively small values of \(p\) (e.g., \(5\) to \(7\)) are used to limit complexity. As a result, the performance degrades, and the candidate set \(\Omega\) may not contain the true ML codeword. Consequently, erroneous decisions \(\bm{d}\) can produce extrinsic information with incorrect signs, leading to error propagation during iterative decoding.

Low-complexity threshold-based methods can detect such errors. For example, \cite{ArtemasovTransformer} proposes two criteria inspired by \cite{Forney}. The first, referred to as the top-1 method, marks \(\bm{d}\) as unreliable if \(\|\bm{r} - \mu(\bm{d})\|^2 > T_1\), where \(T_1\) is an optimizable threshold. The second, known as the top-2 method, additionally considers the second most likely codeword \(\bm{d}' =\arg\min_{\bm{c}_i\in\Omega,\bm{c}_i\neq \bm{d}}\|\bm{r}-\mu(\bm{c}_i)\|^2\). The decision \(\bm{d}\) is then marked as unreliable if
\begin{equation}
    \label{eq:top2}
    \|\bm{r} - \mu(\bm{d}')\|^2 - \|\bm{r} - \mu(\bm{d})\|^2 < T_2,
\end{equation}
where \(T_2\) is another optimizable threshold. Alternatively, anchor-based approaches, such as in \cite{miao2022JLT}, can be employed.

These methods introduce negligible additional complexity but offer limited performance gains, as they rely on a single decision criterion. In particular, achieving a high detection rate of erroneous decisions often comes at the cost of a high false-alarm rate. To overcome this limitation, \cite{ArtemasovTransformer} proposes a transformer-based model that takes the soft information \(\bm{r}\) and the candidate set \(\Omega\) as input to predict erroneous decisions more accurately. While this approach yields significant performance improvements, it also incurs substantial computational overhead, since the model must be evaluated for each component code decoding.

\vspace{-0.5em}
\section{Proposed Decoding Algorithm}
\begin{figure}
    \centering
    \begin{tikzpicture}[
    x=1cm, y=1cm,
    neuron/.style={circle, fill=black, inner sep=1.6pt},
    hiddenneuron/.style={draw = black, fill=white, circle, inner sep=6pt},
    every node/.style={font=\small}
]

\foreach \i in {1,...,10} {
    \node[neuron] (I\i) at (2,5.5-\i*0.25) {};
    \node[left] at (I\i.west) {$x_{\i}$};
}

\node[hiddenneuron] (O) at (4,4) {};

\foreach \j in {1,...,10} {
    \draw[thin] (I\j) -- (O);
}

\node at (4,4) {$\sigma(\cdot)$};

\draw[-latex] (O.east) -- +(0.6,0) node[right] {$\hat{y}$};

\end{tikzpicture}
    \caption{Proposed regression model for detecting erroneous Chase decoding decision.}
    \label{fig:nn}
\end{figure}
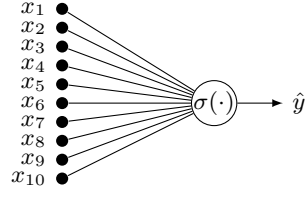

We propose to estimate the confidence of the Chase decoding decision using a simple regression model with 10 inputs and a single output neuron as depicted in Fig.~\ref{fig:nn}. The input features include the noise standard deviation $\sigma$, the ratio of unique candidate codewords to test patterns $|\Omega|/2^p$, the normalized Euclidean distances $d^{\text{E}}_i/n$ of the four best candidate codewords, and their corresponding destructive Euclidean distances
\[
\sum_{1 \leq j \leq n,\, c_{i,j} \neq \psi(r_j)} (r_j - \mu(c_{i,j}))^2,
\]
as defined in \cite{NongDistance}. These features are designed to efficiently capture the reliability of the Chase decision.

The features $\bm{x} \in \mathbb{R}^{10}$ are linearly combined with trainable weights $\bm{w} \in \mathbb{R}^{10}$ and bias $\epsilon$, followed by a sigmoid activation $\sigma(x)=\frac{1}{1+e^{-x}}$, yielding
\[
\hat{P} = \sigma(\bm{w}^{\textsf{T}}\bm{x} + \epsilon).
\]
The output \(\hat{P}\) represents the estimated probability that the Chase decision is erroneous.

We optimize the model parameters offline via supervised learning using batches of \(1280\) samples. The label \(P\) denotes the ground truth, where \(P=0\) indicates \(\bm{d}=\bm{b}\) is correct and \(P=1\) an erroneous estimate. Training is independent of specific codewords and relies only on the selected features. We use binary cross-entropy loss \cite[Chap. 4]{bishop2006prml} and train for \(5000\) epochs with a learning rate of \(0.001\).

The normalized confusion matrix of the trained neural network is shown in Tab.~\ref{tab:table1}. Its prediction accuracy significantly outperforms simpler criteria, which exhibit false-alarm rates of nearly \(0.5\) when tuned to achieve the same correct-marking rate in our experiments.

\begin{table}[t!]
    \centering
    \caption{Normalized confusion matrix of the prediction of the trained neural network}
    \label{tab:table1}
    \resizebox{0.7\linewidth}{!}{
    \begin{tabular}{ccc}
    \toprule
        & Erroneous $\bm{d}$ & Correct $\bm{d}$ \\ 
        \midrule
        Flagged & $0.94$ & $0.32$ \\
        Not flagged & $0.06$ & $0.68$ \\
    \bottomrule
    \end{tabular}
    }
\end{table}

The trained model is then used to assist Chase--Pyndiah decoding by introducing a small modification. After performing Chase decoding and obtaining set $\Omega$ and decision $\bm{d}$, we compute the selected features and evaluate the output of the neural network. Note that we can skip the sigmoid function evaluation. If $\bm{w}\bm{x}^{\mathsf{T}}+b>0$, the decision $\bm{d}$ is flagged as unreliable.  Then, when computing the extrinsic information, the flagged $\bm{l}_i$ for the rows (columns) are scaled with an additional scaling factor $\gamma$. The proposed decoding algorithm is summarized in Alg.~\ref{algo:decoder}. 

For comparison, we propose an improved variant of the top-2 approach in \cite{ArtemasovTransformer}, termed the \emph{scaled top-2} method, as the top-2 approach performs best among the single-criterion methods. Specifically, \eqref{eq:top2} is used to determine the reliability of the Chase decision \(\bm{d}\), while the remaining decoding procedure follows Alg.~\ref{algo:decoder}. In contrast to the original top-2 approach, which sets the extrinsic information to \(\bm{l}=\bm{0}\) for flagged decisions, the proposed method applies scaling instead.

\vspace{-0.5em}
\section{Numerical Results}
\begin{figure}[t]
    \centering
    \input{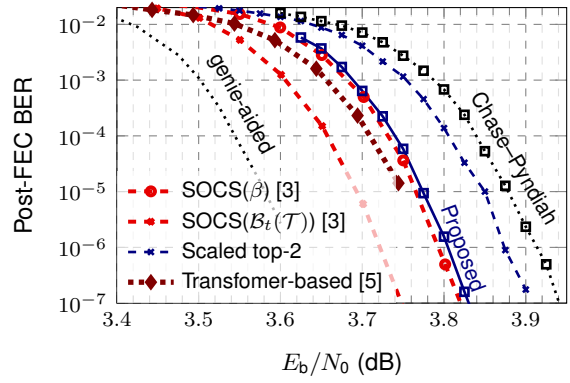}
    \caption{Decoding results of variants of Chase-Pyndiah decoding with $I=4$ and $p=5$.}
    \label{fig:p5}
\end{figure}

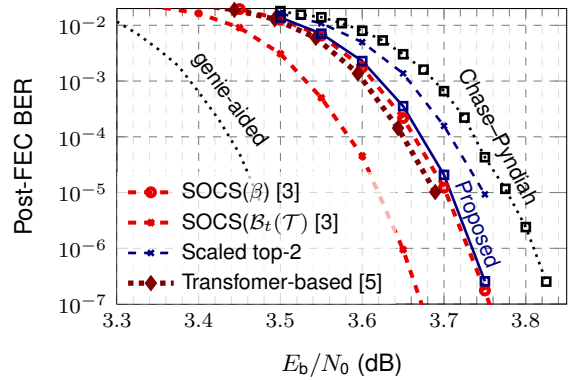
\begin{figure}[t]
    \centering
    \vspace{2em}
    \begin{tikzpicture}
		\pgfplotsset{grid style={dashed, gray}}
		\pgfplotsset{every tick label/.append style={font=\footnotesize}}
		\begin{axis}[%
xmin=3.3,
xmax=3.85,
ymode=log,
ymin=1e-7,
ymax=2e-2,
yminorticks=true,
axis background/.style={fill=white, mark size=1.5pt},
xmajorgrids,
xminorgrids,
ymajorgrids,
yminorgrids,
width=7.5cm,
height=5.5cm,
xtick={3.3, 3.4,...,10.5},
minor x tick num=4,
minor grid style={gray!25},%
ytick={1,0.1,0.01,0.001,1e-4,1e-5,1e-6,1e-7,1e-8,1e-9,1e-10,1e-11,1e-12,1e-13,1e-14,1e-15},
xlabel={$E_{\text{b}}/N_{\text{0}}$ (dB)},
ylabel={Post-FEC BER},
label style={font=\small},
legend cell align={left},
legend style={{at = (0,0)}, {anchor = south west}, draw=none, fill opacity=0.7, text opacity = 1,legend columns=1,font=\footnotesize, row sep = 0pt}
		]

 \addplot [color=red!90!black, dashed, line width=1.5pt, mark=o, mark options={solid,fill=blue!50!black, mark size=1.5pt}]table[x=EbNo,y=BER, col sep=semicolon,row sep=crcr] {
 EbNo;BER\\
3.2; 0.0338982\\
3.25; 0.0316724\\
3.3; 0.0295277\\
3.35; 0.0271138\\
3.4; 0.023662\\
3.45; 0.0195692\\
3.5; 0.0127781\\
3.55; 0.00655514\\
3.6; 0.0017639\\
3.65; 0.000217186\\
3.7; 1.22519e-05\\
3.75; 1.75516e-07\\
3.8; 1.08911e-09\\
 };
\addlegendentry{SOCS($\beta$) \cite{JanzSOCS}}

 \addplot [color=red!90!black, dashed, line width=1.5pt, mark=x, mark options={solid,fill=blue!50!black, mark size=1.5pt}]table[x=EbNo,y=BER, col sep=semicolon,row sep=crcr] {
 EbNo;BER\\
3.2; 0.0298108\\
3.25; 0.0279238\\
3.3; 0.0255475\\
3.35; 0.0214043\\
3.4; 0.0163739\\
3.45; 0.00906313\\
3.5; 0.00309248\\
3.55; 0.000500394\\
3.6; 4.53128e-05\\
3.65; 9.46303e-07\\
3.7; 6.59074e-09\\
 };
\addlegendentry{SOCS($\mathcal{B}_t(\mathcal{T})$ \cite{JanzSOCS}}

\addplot [color=blue!50!black, dashed, line width=1pt, mark=x, mark options={solid,fill=white, mark size=1.5pt}]table[x=EbNo,y=BER, col sep=semicolon,row sep=crcr] {
  EbNo;        EsNo;       delta; ErasureProb;  totalFrame;    FE;         FER;         BER;      SPsize;          BE;  throughput;    mis rate;\\
   3.5;     2.90316;   0.0241068;           0;         392;   391;    0.997449;    0.016273;     1069.19;      418054; 4.46504e+07;           0;\\ %
  3.55;     2.95316;   0.0234675;           0;         392;   371;    0.946429;   0.0108716;     752.811;      279293;  4.4667e+07;           0;\\ %
   3.6;     3.00316;    0.022839;           0;         392;   293;    0.747449;  0.00498523;     437.102;      128071; 4.48772e+07;           0;\\ %
  3.65;     3.05316;   0.0222213;           0;         392;   138;    0.352041;  0.00137504;     255.978;       35325; 4.47692e+07;           0;\\ %
   3.7;     3.10316;   0.0216142;           0;        1568;   118;   0.0752551;  0.00015686;     136.602;       16119; 4.34272e+07;           0;\\ %
  3.75;     3.15316;   0.0210179;           0;       19600;   101;  0.00515306; 9.28217e-06;      118.05;       11923; 4.51001e+07;           0;\\ %
 };
\addlegendentry{Scaled top-2}

 \addplot [color=red!50!black, dotted, line width=2pt, mark=diamond, mark options={solid,fill=blue!50!black, mark size=1.5pt}]table[x=EbNo,y=BER, col sep=semicolon,row sep=crcr] {
 EbNo;BER\\
3.393458092047887;0.024240560910370357\\
3.4436695157400285;0.018905688781804656\\
3.4931084265312277;0.012897462428364311\\
3.543244818618259;0.0059533786397129775\\
3.5947150928809208;0.0013703504519868611\\
3.6437774388039756;0.0001431955996248524\\
3.6892827395393765;0.00001023037892770285\\
 };
\addlegendentry{Transfomer-based \cite{ArtemasovTransformer}}

 \addplot [color=black, dotted, line width=1pt, mark=square, mark options={solid,fill=white, mark size=1.5pt}]table[x=EbNo,y=BER, col sep=semicolon,row sep=crcr] {
  EbNo;        EsNo;       delta; ErasureProb;  totalFrame;    FE;         FER;         BER;      SPsize;          BE;  throughput;    mis rate;\\
      3.5;     2.90316;   0.0241068;           0;         512;   512;           1;   0.0179873;     1178.82;      603554; 4.18502e+07;           0;\\ %
 3.525;     2.92816;   0.0237858;           0;         512;   512;           1;   0.0154992;     1015.76;      520067; 5.23405e+07;           0;\\ %
  3.55;     2.95316;   0.0234675;           0;         512;   511;    0.998047;   0.0139403;     915.382;      467760; 5.32664e+07;           0;\\ %
 3.575;     2.97816;   0.0231519;           0;         512;   503;    0.982422;   0.0109131;         728;      366184; 5.33503e+07;           0;\\ %
   3.6;     3.00316;    0.022839;           0;         512;   479;    0.935547;  0.00797701;     558.797;      267664; 4.49607e+07;           0;\\ %
 3.625;     3.02816;   0.0225288;           0;         512;   421;    0.822266;  0.00535595;     426.879;      179716; 5.30887e+07;           0;\\ %
  3.65;     3.05316;   0.0222213;           0;         512;   328;    0.640625;  0.00302964;     309.933;      101658;  5.2386e+07;           0;\\ %
 3.675;     3.07816;   0.0219164;           0;         512;   233;    0.455078;  0.00160584;     231.258;       53883; 5.01943e+07;           0;\\ %
   3.7;     3.10316;   0.0216142;           0;         512;   121;    0.236328; 0.000659883;     182.992;       22142; 4.39283e+07;           0;\\ %
 3.725;     3.12816;   0.0213147;           0;        1024;   105;    0.102539; 0.000221238;       141.4;       14847; 5.07555e+07;           0;\\ %
  3.75;     3.15316;   0.0210179;           0;        3584;   101;   0.0281808;  4.3026e-05;     100.059;       10106; 4.86954e+07;           0;\\ %
 3.775;     3.17816;   0.0207237;           0;       11776;   102;  0.00866168; 1.16734e-05;     88.3235;        9009; 4.76673e+07;           0;\\ %
  3.8;     3.20316;   0.0204322;           0;       58368;   100;  0.00171327; 2.40249e-06;        91.9;        9190; 4.71648e+07;           0;\\ %
   3.825;     3.22816;   0.0201433;           0;      358912;   100;  0.00027862; 2.52703e-07;       59.44;        5944; 4.71489e+07;           0;\\ %
 }node [pos=0.5,anchor=south,font=\footnotesize,sloped] {Chase--Pyndiah};

 \addplot [color=blue!50!black, line width=1pt, mark=square, mark options={solid,fill=white, mark size=1.5pt}]table[x=EbNo,y=BER, col sep=semicolon,row sep=crcr] {
  EbNo;        EsNo;       delta; ErasureProb;  totalFrame;    FE;         FER;         BER;      SPsize;          BE;  throughput;    mis rate;\\
   3.5;     2.90316;   0.0241068;           0;         512;   502;    0.980469;   0.0138096;     923.052;      463372; 1.10716e+08;   0.0914961;\\ %
  3.55;     2.95316;   0.0234675;           0;         512;   418;    0.816406;   0.0070819;      568.49;      237629;  1.0637e+08;   0.0774031;\\ %
   3.6;     3.00316;    0.022839;           0;         512;   212;    0.414062;  0.00228345;     361.415;       76620; 1.11964e+08;   0.0628443;\\ %
  3.65;     3.05316;   0.0222213;           0;        1024;   102;   0.0996094;  0.00035435;     233.137;       23780; 9.73247e+07;   0.0505757;\\ %
   3.7;     3.10316;   0.0216142;           0;        5120;    52;   0.0101563; 2.07633e-05;     133.981;        6967; 1.03167e+08;   0.0403322;\\ %
  3.75;     3.15316;   0.0210179;           0;      284160;    51; 0.000179476;  2.5587e-07;     93.4314;        4765; 1.03976e+08;   0.0334471;\\ %
 }node [pos=0.8,anchor=south,font=\footnotesize,sloped,yshift=-3pt] {Proposed};

 \addplot [color=black, dotted, line width=1pt, mark=none]table[x=EbNo,y=BER, col sep=semicolon,row sep=crcr] {
  EbNo;        EsNo;       delta; ErasureProb;  totalFrame;    FE;         FER;         BER;      SPsize;          BE;  throughput;    mis rate;\\
   3.3;     2.70316;   0.0267716;           0;         392;   327;    0.834184;   0.0116513;     915.361;      299323; 4.58129e+07;           0;\\ %
 3.325;     2.72816;   0.0264291;           0;         392;   254;    0.647959;   0.0070199;     710.008;      180342; 5.63366e+07;           0;\\ %
  3.35;     2.75316;   0.0260892;           0;         392;   163;    0.415816;  0.00378301;     596.233;       97186;  5.6294e+07;           0;\\ %
 3.375;     2.77816;   0.0257521;           0;         392;   104;    0.265306;  0.00188298;     465.135;       48374; 5.62047e+07;           0;\\ %
   3.4;     2.80316;   0.0254176;           0;        1176;   138;    0.117347; 0.000637249;     355.891;       49113; 5.21369e+07;           0;\\ %
 3.425;     2.82816;   0.0250859;           0;        2352;   111;   0.0471939; 0.000199779;     277.423;       30794; 5.08003e+07;           0;\\ %
  3.45;     2.85316;   0.0247568;           0;        8232;   100;   0.0121477; 4.98469e-05;      268.92;       26892; 5.46039e+07;           0;\\ %
 3.475;     2.87816;   0.0244304;           0;       49000;   100;  0.00204082; 6.93808e-06;       222.8;       22280;  5.2285e+07;           0;\\ %
 }node [pos=0.5,anchor=south,font=\footnotesize,sloped] {genie-aided};

\end{axis}
\end{tikzpicture}
    \caption{Decoding results of variants of Chase-Pyndiah decoding with $I=4$ and $p=6$.}
    \label{fig:p6}
\end{figure}

We consider a PC \(\mathcal{C}_{\text{PC}}\) constructed from the extended BCH component code \(\mathcal{C}_{\text{BCH}}[n=256,k=239,d=6]\), yielding a rate \(R=(k/n)^2 \approx 0.87\).
Decoding results for various schemes with \(I=4\) and \(p=5\) and \(p=6\) are shown in Fig.~\ref{fig:p5} and Fig.~\ref{fig:p6}, respectively.

As a baseline, we use the original Chase--Pyndiah decoder \cite{PyndiahTPC}.
The parameters \(\bm{\alpha}\) and \(\bm{\beta}\) from \cite{PyndiahTPC}, optimized for a different code, perform poorly in this setting.
We therefore optimize the parameters using a Bayesian approach~\cite{optuna_2019}, while constraining $\bm{\alpha}$ and $\bm{\beta}$ to increase with the number of half-iterations. The resulting configuration is used as the baseline. Note that the parameters obtained through this method yield improved performance of the Chase--Pyndiah decoder compared to those reported in~\cite{ArtemasovTransformer} and in ~\cite{JanzSOCS,StrasshoferGMI}, where the optimization is performed greedily using the generalized mutual information.

Next, we evaluate the scaled top-2 approach, which yields approximately $0.03$\,dB of decoding gain compared to the original Chase--Pyndiah decoder at a post-decoding error rate of \(10^{-5}\). In contrast, the original top-2 method in~\cite{ArtemasovTransformer} performs similarly to our Chase--Pyndiah decoder with optimized parameters and is therefore omitted from the figures for clarity.

For completeness, we also include a genie-aided decoder where \(\bm{\alpha}=\bm{\beta}=\bm{1}\) and extrinsic messages are set to \(\bm{0}\) whenever \(\bm{d}\neq \bm{b}\).

The proposed neural network-assisted scaling significantly outperforms both the baseline Chase--Pyndiah decoder and top-2 methods.
Comparisons with \cite{JanzSOCS} and \cite{ArtemasovTransformer} further show that the proposed low-complexity method achieves performance close to the transformer-based and SOCS\((\beta)\) approaches, despite their higher computational complexity.

\vspace{-0.5em}
\section{Complexity Analysis}
We assess the computational overhead relative to the original Chase--Pyndiah decoder.
Per component decoding, the proposed method requires four additional destructive Euclidean distance evaluations, corresponding in the worst case to four sums of \(p+t\) real values.
The neural network evaluation involves only \(10\) multiplications and additions.
The cost of message scaling can be eliminated by precomputing \(\alpha\gamma\).
Overall, the additional complexity is negligible compared to \cite{ArtemasovTransformer} and \cite{JanzSOCS}.
In particular, a single layer of the transformer model in \cite{ArtemasovTransformer} requires \(\mathcal{O}(n^2)\) multiplications.
The SOCS approach in \cite{JanzSOCS} computes extrinsic information by evaluating probabilities over the full candidate set $\Omega$ for each bit, rather than using \eqref{eq:extrinsicInfo}, resulting in significantly higher complexity.

\vspace{-0.5em}
\section{Conclusions}
We present a simple enhancement of the conventional Chase--Pyndiah decoding algorithm, which achieves notable performance gains. All numerical results are produced using the C++/Python implementation available at \url{https://github.com/kit-cel/Chase-Pyndiah-demo}.

\section{Acknowledgement}
This work has received funding from the European Research Council (ERC) under the European Union's Horizon 2020 research and innovation programme (grant agreement No. 101001899). We thank T.~Janz for providing the numerical results of the SOCS decoders.

\printbibliography

\vspace{-4mm}

\end{document}